\documentclass[aps,preprint,showpacs,superscriptaddress,groupedaddress]{revtex4-1} 

\usepackage{verbatim}
\usepackage{multirow}
\usepackage{dcolumn}   
\usepackage{bm}        
\usepackage{amssymb}   
\usepackage{amsmath}
\usepackage{amsfonts}
\usepackage{graphicx}
\usepackage[ngerman, english]{babel}
\usepackage{float}
\usepackage{youngtab}
\usepackage[utf8]{inputenc}
\usepackage{mathrsfs}
\usepackage{rotating}
\usepackage{braket}
\usepackage{subfigure}
\usepackage{color}

\begin{document}
\preprint{DO-TH 16/31}
\title{Discriminating sterile neutrinos and unitarity violation with CP invariants}

\author{Heinrich P\"as}
\email[]{heinrich.paes@tu-dortmund.de}
\author{Philipp Sicking}
\email[corresponding author: ]{philipp.sicking@tu-dortmund.de}
\affiliation{Fakult\"at für Physik,
	Technische Universit\"at Dortmund, 44221 Dortmund,
Germany}

\begin{abstract}
	We present a new method to analyze upcoming results in the search for CP violating neutrino oscillations. The CP violating amplitudes $\mathcal{A}_{\alpha\beta}^{kj}$ provide parametrization independent observables, which will be accessible
	by experiments soon. The strong prediction of a unique $\mathcal{A}_{\alpha\beta}^{kj}$ (the Jarlskog invariant) in case of the standard three neutrino model does not hold in models with new physics beyond the Standard Model. Nevertheless there are still correlations among the amplitudes
	depending on the specific model. Due to these correlations it is possible to reject specific new physics models by determining only 3 of the CP violating amplitudes.
\end{abstract}

\maketitle
\section{Introduction}
The experimental observation of neutrino oscillations and its interpretation as a consequence of neutrino masses provided the first manifestation
of new physics beyond the Standard Model (SM). 
The first conclusive evidence of neutrino oscillation by SNO \cite{Ahmad:2002jz,Barros:2013axa}
and Super-Kamiokande \cite{Fukuda:1998mi} was honored recently by the Nobel Prize of Physics in 2015. With the exception of some anomalies, 
almost all current data can be well explained by a model of three neutrinos with two mass squared differences, $\Delta m_{31}^2$ and $\Delta m_{21}^2$, three mixing
angles $\theta_{12}$, $\theta_{23}$ and  $\theta_{31}$, and one CP phase $\delta$ \cite{Esteban:2016qun}. All parameters are measured to a relatively high precision, except for the octant of $\theta_{23}$,
the mass-ordering and the CP phase. Ongoing and upcoming neutrino experiments 
will narrow down the viable space for these parameters (see \cite{Agashe:2014kda} for a review). A first hint for a maximal $\delta=[-2.03,-0.49](\rm{NH}),\, [-1.87, -0.98](\rm{IH})$ at 90\% CL has been reported by T2K \cite{Abe:2015awa,T2K:Neutrino2016}.

This situation cannot be understood as a proof of the minimal three neutrino picture, though. As has been shown 
by several authors, new physics models can fake a signal at current experiments which look like satisfying the three neutrino paradigm \cite{Goswami:2008mi,Miranda:2016wdr,Ge:2016xya,deGouvea:2016pom,Dutta:2016vcc}.\\
Neutrino oscillation probabilities are described by
introducing the mixing matrix $U$, parametrizing the transformation from neutrino mass to flavor eigenstates, $\ket{\nu_\alpha}=\sum_k U_{\alpha k}\ket{\nu_k}$:
\begin{align}
	P_{\nu_\alpha \rightarrow \nu_\beta}(t)=&\sum_{k,j} U_{\alpha k}^*U_{\beta k} U_{\alpha j} U_{\beta j}^*e^{-i\frac{\Delta m_{kj}^2L}{2E}}\\
	=&\delta_{\alpha\beta} -4\sum_{k>j}^{N} \operatorname{Re}( U_{\alpha k}^*U_{\beta k} U_{\alpha j} U_{\beta j}^*)\sin^2{\left(\frac{\Delta m_{kj}^2L}{4E}\right)}\nonumber \\
	&+2\sum_{k>j}^N\ \operatorname{Im}( U_{\alpha k}^*U_{\beta k} U_{\alpha j} U_{\beta j}^*)\sin{\left(\frac{\Delta m_{kj}^2L}{2E}\right)}\label{eq:Probability_exact},
\end{align}
where ${\mathcal{A}_{\alpha\beta}^{kj}}=\operatorname{Im}( U_{\alpha k}^*U_{\beta k} U_{\alpha j} U_{\beta j}^*)$. 
For antineutrinos the last term switches its sign, so the CP violation $P_{\nu_\alpha \rightarrow \nu_\beta}-P_{\bar{\nu}_\alpha \rightarrow \bar{\nu}_\beta}$ depends only on the CP violating amplitudes $\mathcal{A}_{\alpha\beta}^{kj}$. Here, $N$ indicates the number of light neutrinos involved in the oscillation
process. 
If all neutrino mass eigenstates involved in the oscillation process are light compared to the beam energy $E$, the mixing matrix $U$ is unitary. If, on the other hand, heavy flavors are integrated out, the resulting effective mixing matrix $U$ can be non-unitary. 
Note that in this case in addition to neutrino oscillations  zero-distance-effects can arise, which we do not consider in this work.\\
A common approximative parametrization used in the literature is based on a series expansion in $\alpha=\frac{\Delta m_{31}^2}{\Delta m_{21}^2}$:
\begin{align}
	P_{\nu_e\rightarrow\nu_\mu}\sim A\sin^2{\Delta}+\alpha\sin{\delta}B\sin^{3}{\Delta}+\alpha\cos{\delta}C\cos{\Delta}\sin^2{\Delta}+\alpha^2 D\sin^2{\Delta}
	\label{}
\end{align}
with $\Delta=\frac{\Delta m_{31}L}{4E}$ and $A,B,C,D$ are functions of the standard mixing angles \cite{Freund:2001pn}.\\
The CP violating term (proportional to $\sin{\delta}$) is suppressed by $\alpha$ but the unitarity of $U^{3\times3}$ is implicitly used to derive this formula.
Various efforts exist in the literature to improve the above approximation for new, more exact or shorter parametrizations \cite{GonzalezGarcia:2001mp,Gupta:2001ia,Guo:2001yt,Blasone:2002jv,Rodejohann:2003sc,Li:2004nh,Dattoli:2008zz,Rodejohann:2011vc,Huang:2011by}
or to include matter effects \cite{Kimura:2002hb,Akhmedov:2004ny,Zhou:2011xm,Flores:2015mah,Denton:2016wmg,Johnson:2016tzb,Johnson:2015psa,Flores:2016moy,Minakata:2015gra}.\\
Here we rely on the exact expressions given in equation \eqref{eq:Probability_exact} instead, which is invariant under reparametrization. 
In particular the CP violating amplitudes $\mathcal{A}_{\alpha\beta}^{kj}$ are independent of the 
parametrization \cite{Wagner:1998mc,Weiler:1998uk} and can be determined in various extensions to the SM case. 
A specific feature which had already been pointed out by 
Jarlskog \cite{PhysRevLett.55.1039},\cite{PhysRevD.36.2128} is that in the case of exactly three flavors and a unitary mixing matrix $U$,  all CP violating 
amplitudes $\mathcal{A}_{\alpha\beta}^{kj}$ have identical absolute values. 
This observation was first exploited in the quark sector 
where the famous CKM unitarity triangle provides a
precise test for unitarity and therefore for the SM itself.
Analyses of the lepton sector in terms of unitarity triangles have been worked out in \cite{Xing:2005gk,Bjorken:2005rm,Xing:2012ej,He:2016dco}, but
the insights are limited in cases where the triangle does not close, since the source of unitary violation cannot be determined. \\
Inspired by previous work \cite{Wagner:1998mc,Weiler:1998uk} we take a closer look to sums and ratios of the CP violating amplitudes 
$\mathcal{A}_{\alpha\beta}^{kj}$ and find useful correlations among them. These correlations depend highly on the specific model and therefore provide a useful test for new physics
in CP violating neutrino oscillations.

\section{Analytic treatment of 3+1 $\nu$}\label{sec:analytic}
A popular extension of the three neutrino model is to add an additional light sterile neutrino \cite{Kopp:2013vaa,Abazajian:2012ys}. This is motivated by the LSND \cite{Aguilar:2001ty}, MiniBooNE \cite{AguilarArevalo:2010wv}, 
reactor \cite{Mention:2011rk} and gallium anomalies \cite{Giunti:2010zu} but in conflict with a recent IceCube analysis \cite{TheIceCube:2016oqi}.
In this model the mixing matrix $U$ is now a $4\times 4$ unitary mixing matrix but the $3\times 3$ sub matrix is not unitary anymore. Although the resulting
amplitudes are no longer unique, they are related due to the unitarity of the complete mixing matrix. By exploiting these relations in the context of the
quark sector it has been shown for four flavors
that all amplitudes can be reduced to only three independent CP violating amplitudes \cite{2009JMP....50l3526S}.
In the following we follow these arguments translated to the notation commonly used in neutrino physics.
\\
In total there exist $4 \times 4\times 4\times 4=256$ ($\alpha, \beta \in \{e,\mu,\tau,s\}$ and $ k,j \in \{1,2,3,4\}$) 
different CP violating amplitudes $\mathcal{A}_{\alpha\beta}^{kj}=\operatorname{Im}( U_{\alpha k}^*U_{\beta k} U_{\alpha j} U_{\beta j}^*)$, whereas the number is strongly reduced by the fact that $\mathcal{A}_{\alpha\beta}^{kj}=0 \text{ for } 
\alpha=\beta \text{ or } k=j$ and due to symmetry, $\mathcal{A}_{\alpha\beta}^{kj}=\mathcal{A}_{\beta\alpha}^{kj}$ and $\mathcal{A}_{\alpha\beta}^{kj}=\mathcal{A}_{\alpha\beta}^{jk}$
Therefore it is sufficient to only consider 
$\mathcal{A}_{\alpha\beta}^{kj}$ where $\alpha<\beta$ and $k>j$.
Note that the previous relations hold due to the definition of $\mathcal{A}_{\alpha\beta}^{kj}$ regardless of the underlying $U$ and are not specific for the 3+1$\nu$ model.
This reduces the number of CP violating amplitudes to 36. These 36 amplitudes are not independent of each other and can be expressed via only nine amplitudes (see Appendix \ref{App_Relations}).
Again, these nine amplitudes can be expressed by three remaining amplitudes via the following expression
\begin{align}
	\begin{pmatrix}
		\mathcal{A}_{e\mu}^{32}\\
		\mathcal{A}_{e\mu}^{43}\\
		\mathcal{A}_{\mu\tau}^{21}\\
		\mathcal{A}_{\mu\tau}^{43}\\
		\mathcal{A}_{\tau s}^{21}\\
		\mathcal{A}_{\tau s}^{32}\\
	\end{pmatrix}
	= \bf{M}^{-1}\begin{pmatrix}
		\mathcal{R}_{e\mu}^{32}\mathcal{A}_{e\mu}^{21}\\
		\mathcal{R}_{\mu\tau}^{43}\mathcal{A}_{\tau s}^{43}\\
		\mathcal{R}_{\mu\tau}^{21}\mathcal{A}_{e\mu}^{21}\\
		\mathcal{R}_{\tau s}^{32}\mathcal{A}_{\tau s}^{43}\\
		(\mathcal{R}_{\tau\tau}^{32}+\mathcal{R}_{e\mu}^{32})\mathcal{A}_{\mu\tau}^{32}\\
		(\mathcal{R}_{\mu\tau}^{33}+\mathcal{R}_{e\mu}^{32})\mathcal{A}_{\mu\tau}^{32}\\
	\end{pmatrix}
	\label{eq_master}\text{,}
\end{align}
with $\bf{M}^{-1}$ defined by the inverse of
\begin{align}
	\bf{M}=\begin{pmatrix}
		-(\mathcal{R}_{e\mu}^{22}+\mathcal{R}_{e\mu}^{21})& \mathcal{R}_{e\mu}^{22} & 0 & 0 & 0 & 0\\
		0 & \mathcal{R}_{\tau\tau}^{43} & 0 & -(\mathcal{R}_{\tau\tau}^{43}+\mathcal{R}_{\tau s}^{43}) & 0 & 0\\
		0 & 0 & -(\mathcal{R}_{\mu\mu}^{21}+\mathcal{R}_{e\mu}^{21}) & 0 & \mathcal{R}_{\mu\mu}^{21} & 0\\
		0 & 0 & 0 & 0 & \mathcal{R}_{\tau s}^{33} & -(\mathcal{R}_{\tau s}^{33}+\mathcal{R}_{\tau s}^{43}) \\
		\mathcal{R}_{\tau \tau}^{32} & 0 & 0 & 0 & 0 & -\mathcal{R}_{\mu\tau}^{32}\\
		0 & 0 & -\mathcal{R}_{\mu\tau}^{33} & -\mathcal{R}_{\mu\tau}^{32} & 0 & 0
	\end{pmatrix}\text{.}
\end{align}
The amplitudes $\mathcal{R}_{\alpha\beta}^{kj}=\operatorname{Re}( U_{\alpha k}^*U_{\beta k} U_{\alpha j} U_{\beta j}^*)$ correspond to the CP conserving amplitudes in neutrino oscillations. These relations therefore provide a connection
between the CP violating and the CP conserving processes.\\
To emphasize the differences between $3\nu$ and $3+1\nu$ we want to highlight following relations:
\begin{align}
	\mathcal{A}_{e\mu}^{31}&=-\mathcal{A}_{e\mu}^{32}+\mathcal{A}_{e\mu}^{43}\\
	\mathcal{A}_{e\tau}^{21}&=-\mathcal{A}_{\mu\tau}^{32}+\mathcal{A}_{\tau s}^{43}\\
	\mathcal{A}_{e\tau}^{31}&=-\mathcal{A}_{e\tau}^{32}-\mathcal{A}_{\tau s}^{32}+\mathcal{A}_{\tau s}^{43}
	\label{eq:relations}
\end{align}
The relations reduce to the $3\nu$ case, if no mixing with the light neutrino takes place. This corresponds to
vanishing non diagonal elements in the fourth line and column of $U$. Consequently, all amplitudes vanish if $\alpha \lor \beta = s$ or $k \lor j = 4$. Due to the expected smallness of mixing with sterile states, the deviations 
from uniform amplitudes in the $3\times 3$ sector could be treated in a perturbation approach.

\section{Numeric Analysis of sterile neutrinos and non-unitary scenarios}
The relations in the previous section rely on the unitarity of the resulting $3+1 \nu$ model. In general these relations are, if possible, harder to find and more
complicated. An easier approach is to use a numeric analysis of the correlations of the different amplitudes for different models. 
Therefore we pick random numbers for all parameters in the specific model (SM and BSM parameters) and generate the resulting mixing matrix $U$. To check if the generated
combination of parameters satisfy current experimental bounds, we compare the entries of the $3\times 3$ sub matrix of $U$ with the bounds presented in \cite{Parke:2015goa}, 
where a global fit is performed without implying a unitarity of $U^{3\times 3}$.
\begin{align}
	|U|_{3\sigma}^{3\times3}=\begin{pmatrix}
		0.76\rightarrow 0.85 & 0.50 \rightarrow 0.60 & 0.13 \rightarrow 0.16\\
		0.21\rightarrow 0.54 & 0.42 \rightarrow 0.70 & 0.61 \rightarrow 0.79\\
		0.18\rightarrow 0.58 & 0.38 \rightarrow 0.72 & 0.40 \rightarrow 0.78
	\end{pmatrix}\label{eq_Ubounds}                                                                                                     
\end{align}
For a viable combination of parameters all accessible amplitudes $\mathcal{A}_{\alpha\beta}^{kj}$ are calculated and extracted. For each model we extracted 100,000 viable
combinations. To show the correlation we 
performed a kernel density estimation for different combination of amplitudes, i.e. estimating the underlying probability density function by summing up Gaussian kernels placed on every data point.\\
We compare 4 different approaches of neutrino physics beyond the three neutrino paradigm:
\begin{enumerate}
	\item[(i)] a model of one additional light sterile neutrino ($3+1\nu$), motivated by LSND \cite{Aguilar:2001ty}, MiniBooNE- \cite{AguilarArevalo:2010wv}, gallium- \cite{Giunti:2010zu} and reactor anomaly \cite{Mention:2011rk}.
		Typically the additional mass squared difference lies in the $\sim 1$ eV range \cite{Kopp:2013vaa,Abazajian:2012ys}. Due to the low mass the sterile state
		participates in the oscillation. The sterile neutrino does not interact via SM gauge interactions with other SM particles. The mixing matrix is a $4\times 4$ unitary matrix (see sec. \ref{sec:analytic} for more details).
	\item[(ii)] a model of two additional light sterile neutrinos ($3+2\nu$), similar to model (i) but with an extended parameter space (additional mixing angles and CP phases) due to the additional sterile state. The mixing matrix is a $5\times 5$ unitary matrix.
	\item[(iii)] a scenario of non-unitarity without additional constraints (NU). This scenario is realized by   
		modifying the 
		unitary matrix with a lower triangular matrix $\alpha$
		\begin{align}
			U_{NU}=(I-\alpha)U^{3\times 3}=\begin{pmatrix}
				1-\alpha_{ee} & 0 & 0 \\ \alpha_{e \mu} & 1- \alpha_{\mu \mu} & 0\\ \alpha_{\tau e} & \alpha_{\mu\tau} & 1-\alpha_{\tau\tau}
			\end{pmatrix}U^{3\times 3}
			\label{}
		\end{align}
		where $|\eta_{\alpha\beta}|<1$. The diagonal entries are real and the off-diagonal entries are complex parameters (see for instance \cite{Escrihuela:2015wra, FernandezMartinez:2007ms, Antusch:2006vwa}).
	\item[(iv)] a scenario of non-unitarity where additional fermions trigger 
		rare decays like $\mu\rightarrow e\gamma$.
		The corresponding constraints from rare decays and electroweak
		precision observables are
		presented in \cite{Blennow:2016jkn}  ("minimal flavor violation" MUV, the non unitarity is parametrized as in scenario (iii))
		\begin{align}
			\alpha_{ee} <& 1.3\cdot 10^{-3}, & |\alpha_{\mu e}| <& 6.8 \cdot 10^{-4} \nonumber, \\
			\alpha_{\mu\mu} <& 2.0\cdot 10^{-4}, & |\alpha_{\tau e}| <& 2.7 \cdot 10^{-3}, \label{eq_alphaBounds}\\
			\alpha_{\tau\tau} <& 2.8 \cdot 10^{-3}, & |\alpha_{\tau \mu}| <& 1.2 \cdot 10^{-3}. \nonumber
		\end{align}
		These constraints are used as priors in our numeric analysis. Many new physics models can influence neutrino oscillation in a way described by NU and MUV. For instance heavy
		right handed neutrinos introduced in seesaw models or non standard neutrino interaction (NSI) at production and detection
		can be described by the MUV and NU scenarios, respectively.
\end{enumerate}

\section{Results}
\begin{figure}[H]
	\centering
	\includegraphics[width=0.9\textwidth]{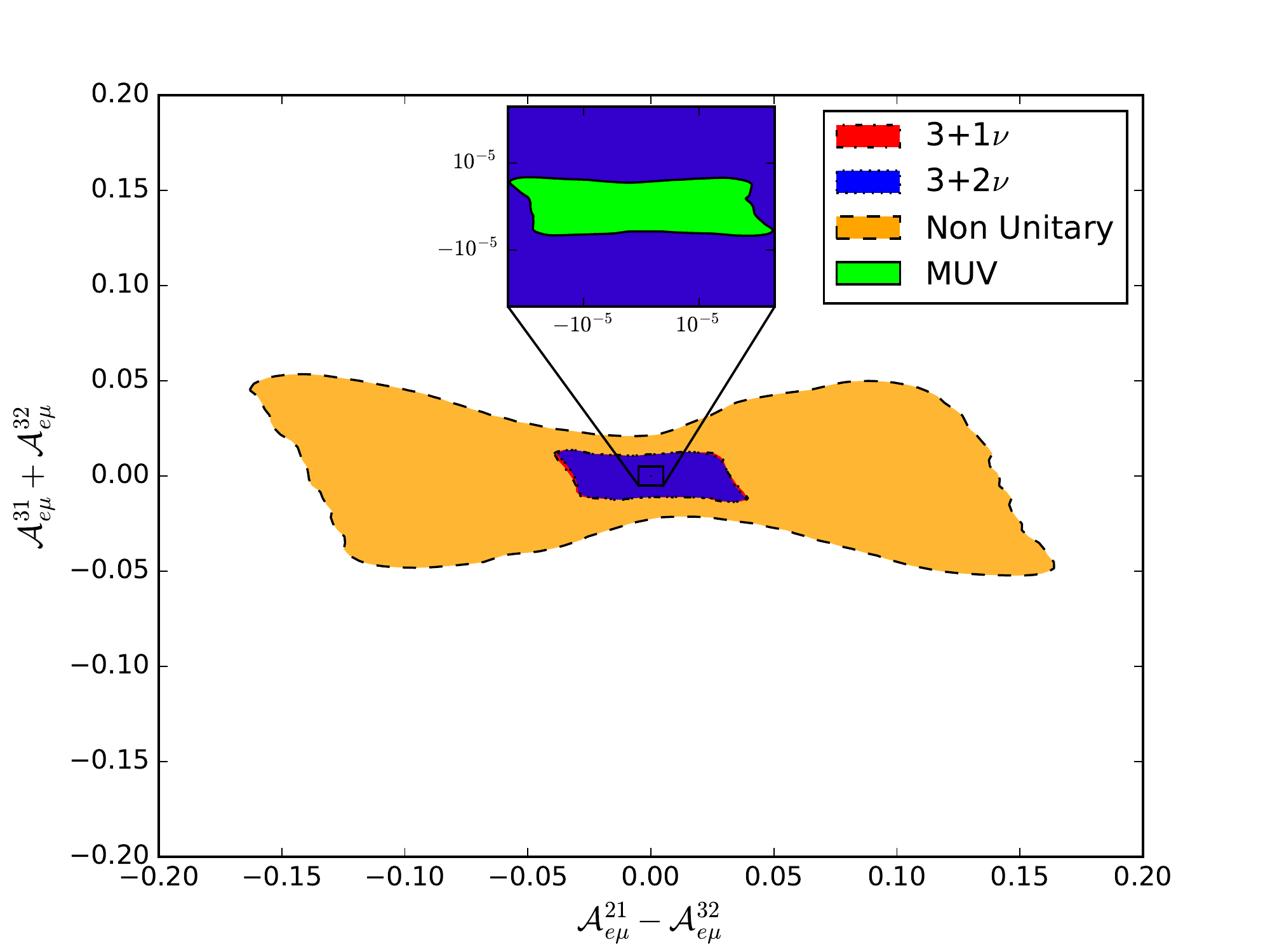}
	\caption{Kernel density estimates for the different scenarios: $3+1\nu$ in red, $3+2\nu$ in blue, Non-Unitarity in yellow and Minimal Unitarity Violation in green.
		Shown is the differences of the 3 different CP 
		violating amplitudes in the $\nu_e\rightarrow\nu_\mu$-channel. The colored area corresponds to the 95\% CL of the KDE.
		The three neutrino prediction corresponds to the point at $(0,0)$. Except for numerical effects, the areas for the $3+1\nu$ and the $3+2\nu$ model match each other.
		A significant deviation between NU and new sterile states can be observed. Due to the strong constraints for MUV, the viable regions are extremely small and deviations from three neutrino prediction
		will be hard to measure.}
	\label{fig:AemuSub}
\end{figure}

\begin{figure}[h]
	\centering
	\includegraphics[width=0.9\textwidth]{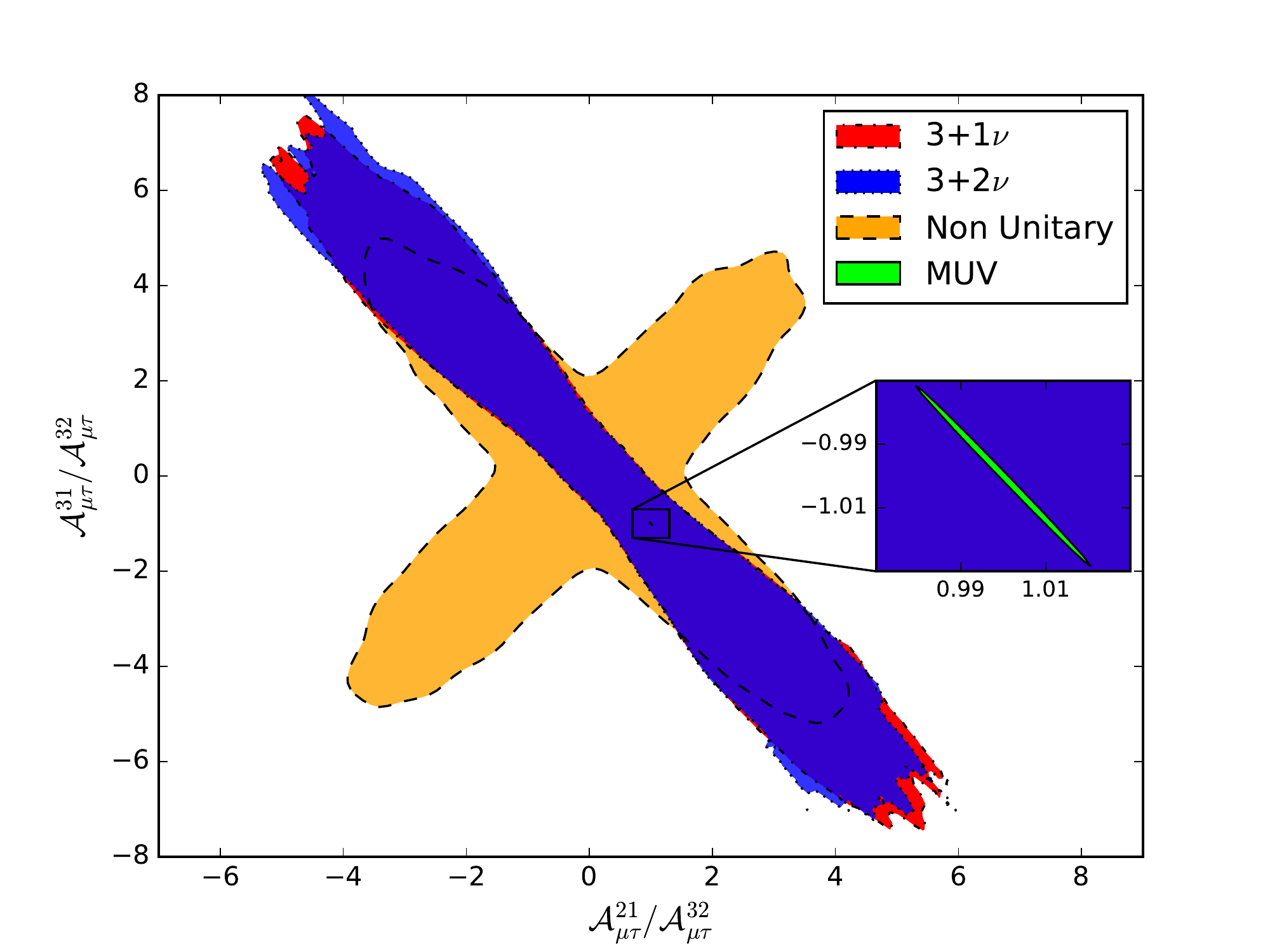}
	\caption{Kernel density estimates for the different scenarios: $3+1\nu$ in red, $3+2\nu$ in blue, Non-Unitarity in yellow and Minimal Unitarity Violation in green.
		Shown is the ratios of the 3 different CP violating amplitudes in the $\nu_\mu\rightarrow\nu_\tau$-channel.
		The colored area corresponds to the 95\% CL of the KDE.
		The three neutrino prediction corresponds to the point at $(1,-1)$. Except for numerical effects, the areas for the $3+1\nu$ and the $3+2\nu$ model match each other.
		A significant deviation between NU and new sterile states can be observed. Due to the strong constraints for MUV, the viable regions are extremely small and deviations from three neutrino prediction
		will be hard to measure.}
	\label{fig:AmutauDiv}
\end{figure}

The 95\% CL of the generated kernel density estimates for oscillations of $\nu_\mu$ are shown in figures \ref{fig:AemuSub} and \ref{fig:AmutauDiv}. We focus on these modes since the production of $\nu_\mu$ is well understood
and the modes are investigated by several current experiments. We do not consider amplitudes where sterile states are involved due to missing detection mechanisms. We also do not consider amplitudes with additional
mass differences beyond the solar and atmospheric $\Delta m^2_{12}$ and $\Delta m^2_{23}$ since these are by now not known and current experiments are optimized for the known mass squared differences.
As can be seen clearly for the scenarios with additional light neutrinos and non unitarity without constraints the corresponding parameter spaces allow for significant deviation from the SM prediction
of uniform CP violating. The MUV scenario albeit provides only a comparatively small allowed region. The strong constraints for the unitary violating parameters $\alpha$ (see equation \eqref{eq_alphaBounds}) as priors 
strongly restrict deviations from the SM prediction. The allowed regions fulfill all current bounds and display the uncertainties in equation \eqref{eq_Ubounds} and the not yet determined CP phase(s).\\
The differences between the $3+1\nu$- and $3+2\nu$-model are negligible. Due to invariance under re-parametrization the amplitudes in the $3\times 3$ sub matrix do not change
by rotations in the 4-5-Plane in case of a $3+2\nu$-model. To investigate a difference between $3+1\nu$ and $3+2\nu$ scenarios, 
amplitudes with sterile states or additional mass squared differences 
have to be taken into account which are not expected to be accessible experimentally in the near future.\\
Comparing the models with additional light neutrinos with the scenario of unconstrained non-unitarity one can find large deviations. The scenario of non unitarity 
provides viable parameter sets which are far outside the 95\% CL of the models with additional light neutrinos.\\
The MUV scenario provides only a small deviation from the SM due to the strong constraints from electroweak precision observables. The expected deviations are out of reach of current experiments. Therefore a sizable
measured deviation from the SM has to have another source than the MUV scenario.\\
Hence the experimental measurement of the corresponding CP violating amplitudes can be a direct test for the three neutrino paradigm and can also discriminate
between different SM extensions: If the experimental values will turn out to lie outside a viable region of $3+1\nu$, $3+2\nu$ or the MUV scenario these models can be ruled out consistently.\\
Similar plots have been fabricated for all combinations of amplitudes and yield similar results. Whether the best discriminators
are provided by the sums or the ratios of amplitudes 
will turn out once experimental data will be available. 
\clearpage
\section{Conclusion}
In this work we have developed a new method to test and discriminate the standard three neutrino paradigm and several extensions based on
the study of various combinations of CP violating amplitudes 
$\mathcal{A}_{\alpha\beta}^{kj}$. These amplitudes are easily accessible via oscillation experiments searching for CP violation. 
The amplitudes
and the relations among them have been translated into the notation commonly used in the neutrino community.
Moreover, the concept has been generalized to scenarios with five neutrinos and non-unitary mixing matrices.
Powerful discriminators between different scenarios of physics beyond the SM can be exploited
once experiments determine three different amplitudes. In this case it is 
possible to rule out not only the three neutrino paradigm but also models of additional sterile light neutrinos or the scenario of MUV in large regions of the respective parameter spaces.
On the other hand, a determination of a
unique amplitude would be in agreement with both the three neutrino model but also with specific parameter combinations of new physics models.
Note, that these calculations rely on the vacuum values of neutrino properties. 
They are independent of specific mass differences. Matter effects are not included yet
but will be addressed in future work. Thus a comparison between different experimental results involving matter effects requires some care. 
\section{Acknowledgement}
We thank Dennis Loose for useful discussion regarding the numerical analysis and Mariam T\'{o}rtola and Matthias Blennow for useful comments.
This project was supported in part by a Fulbright Program grant sponsored by the Bureau of Educational and Cultural Affairs of the United States
Department of State and administered by the Institute of International Education. PS also thanks Vanderbilt University, Prof. Thomas Kephart and Prof. Thomas J. Weiler for the kind hospitality.\clearpage
\begin{appendix}
	\section{Analytic relations of CP violating amplitudes}\label{App_Relations}
	\begin{table}[H]
			\centering
			\begin{tabular}{ll}
			$\mathcal{A}_{e\mu}^{21}=  \mathcal{A}_{e\mu}^{21}$,&
			$\mathcal{A}_{e\mu}^{31}= -\mathcal{A}_{e\mu}^{32}+\mathcal{A}_{e\mu}^{43}$,\\
			$\mathcal{A}_{e\mu}^{41}= -\mathcal{A}_{e\mu}^{21}+\mathcal{A}_{e\mu}^{32}-\mathcal{A}_{e\mu}^{43}$,&
			$\mathcal{A}_{e\mu}^{32}= \mathcal{A}_{e\mu}^{32} $,\\
			$\mathcal{A}_{e\mu}^{42}= \mathcal{A}_{e\mu}^{21}-\mathcal{A}_{e\mu}^{32}$,&
			$\mathcal{A}_{e\mu}^{43}= \mathcal{A}_{e\mu}^{43}$,\\
			$\mathcal{A}_{e\tau}^{21}= -\mathcal{A}_{\mu\tau}^{21}+\mathcal{A}_{\tau s}^{21}$,&
			$\mathcal{A}_{e\tau}^{31}= \mathcal{A}_{\mu\tau}^{32}-\mathcal{A}_{\mu\tau}^{43}-\mathcal{A}_{\tau s}^{32}+ \mathcal{A}_{\tau s}^{43}$,\\
			$\mathcal{A}_{e\tau}^{41}= \mathcal{A}_{\mu\tau}^{21}-\mathcal{A}_{\mu\tau}^{32}+\mathcal{A}_{\mu\tau}^{43}-\mathcal{A}_{\tau s}^{21}+\mathcal{A}_{\tau s}^{32}-\mathcal{A}_{\tau s}^{43}$,\,\,\, &
			$\mathcal{A}_{e\tau}^{32}= -\mathcal{A}_{\mu\tau}^{32}+\mathcal{A}_{\tau s}^{32}$,\\
			$\mathcal{A}_{e\tau}^{42}= -\mathcal{A}_{\mu\tau}^{21}+\mathcal{A}_{\mu\tau}^{32}+\mathcal{A}_{\tau s}^{21}-\mathcal{A}_{\tau s}^{32}$,&
			$\mathcal{A}_{e\tau}^{43}= -\mathcal{A}_{\mu\tau}^{43}+\mathcal{A}_{\tau s}^{43}$,\\
			$\mathcal{A}_{es}^{21}= -\mathcal{A}_{e\mu}^{21}+\mathcal{A}_{\mu\tau}^{21}-\mathcal{A}_{\tau s}^{21} $,&
			$\mathcal{A}_{es}^{31}= \mathcal{A}_{e\mu}^{32}-\mathcal{A}_{e\mu}^{43}-\mathcal{A}_{\mu\tau}^{32}+\mathcal{A}_{\mu\tau}^{43}+\mathcal{A}_{\tau s}^{32}-\mathcal{A}_{\tau s}^{43}$,\\
			\multicolumn{2}{l}{$\mathcal{A}_{es}^{41}= \mathcal{A}_{e\mu}^{21}-\mathcal{A}_{e\mu}^{32}+\mathcal{A}_{e\mu}^{43}-\mathcal{A}_{\mu\tau}^{21}+\mathcal{A}_{\mu\tau}^{32}-\mathcal{A}_{\mu\tau}^{43}+\mathcal{A}_{\tau s}^{21}-\mathcal{A}_{\tau s}^{32}+\mathcal{A}_{\tau s}^{43}$,}\\
			$\mathcal{A}_{es}^{32}= -\mathcal{A}_{e\mu}^{32}+\mathcal{A}_{\mu\tau}^{32}-\mathcal{A}_{\tau s}^{32}$,&
			$\mathcal{A}_{es}^{42}= -\mathcal{A}_{e\mu}^{21}+\mathcal{A}_{e\mu}^{32}+\mathcal{A}_{\mu\tau}^{21}-\mathcal{A}_{\mu\tau}^{32}-\mathcal{A}_{\tau s}^{21}+\mathcal{A}_{\tau s}^{32}$,\\
			$\mathcal{A}_{es}^{43}= -\mathcal{A}_{e\tau}^{43}+\mathcal{A}_{\mu\tau}^{43}-\mathcal{A}_{\tau s}^{43}$,&
			$\mathcal{A}_{\mu\tau}^{21}=  \mathcal{A}_{\mu\tau}^{21}$,\\
			$\mathcal{A}_{\mu\tau}^{31}= -\mathcal{A}_{\mu\tau}^{32}+\mathcal{A}_{\mu\tau}^{43}$,&
			$\mathcal{A}_{\mu\tau}^{41}= -\mathcal{A}_{\mu\tau}^{21}+\mathcal{A}_{\mu\tau}^{32}-\mathcal{A}_{\mu\tau}^{43}$,\\
			$\mathcal{A}_{\mu\tau}^{32}=  \mathcal{A}_{\mu\tau}^{32}$,&
			$\mathcal{A}_{\mu\tau}^{42}= \mathcal{A}_{\mu\tau}^{21}-\mathcal{A}_{\mu\tau}^{32}$,\\
			$\mathcal{A}_{\mu\tau}^{43}= \mathcal{A}_{\mu\tau}^{43}$,&
			$\mathcal{A}_{\mu s}^{21}= \mathcal{A}_{e\mu}^{21}-\mathcal{A}_{\mu\tau}^{21}$,\\
			$\mathcal{A}_{\mu s}^{31}= - \mathcal{A}_{e\mu}^{32}+\mathcal{A}_{e\mu}^{43}+\mathcal{A}_{\mu\tau}^{32}-\mathcal{A}_{\mu\tau}^{43}$,&
			$\mathcal{A}_{\mu s}^{41}= - \mathcal{A}_{e\mu}^{21}+ \mathcal{A}_{e\mu}^{32}- \mathcal{A}_{e\mu}^{43}+\mathcal{A}_{\mu\tau}^{21}- \mathcal{A}_{\mu\tau}^{32}+ \mathcal{A}_{\mu\tau}^{43}$,\\
			$\mathcal{A}_{\mu s}^{32}= \mathcal{A}_{e\mu}^{32}-\mathcal{A}_{\mu\tau}^{32}$,&
			$\mathcal{A}_{\mu s}^{42}= \mathcal{A}_{e\mu}^{21}-\mathcal{A}_{e\mu}^{32}-\mathcal{A}_{\mu\tau}^{21}+\mathcal{A}_{\mu\tau}^{32}$,\\
			$\mathcal{A}_{\mu s}^{43}= \mathcal{A}_{e\mu}^{43}-\mathcal{A}_{\mu\tau}^{43}$,&
			$\mathcal{A}_{\tau s}^{21}=  \mathcal{A}_{\tau s}^{21}$,\\
			$\mathcal{A}_{\tau s}^{31}= -\mathcal{A}_{\tau s}^{32}+\mathcal{A}_{\tau s}^{43}$,&
			$\mathcal{A}_{\tau s}^{41}= -\mathcal{A}_{\tau s}^{21}+\mathcal{A}_{\tau s}^{32}-\mathcal{A}_{\tau s}^{43}$,\\
			$\mathcal{A}_{\tau s}^{32}=  \mathcal{A}_{\tau s}^{32}$,&
			$\mathcal{A}_{\tau s}^{42}= \mathcal{A}_{\tau s}^{21}-\mathcal{A}_{\tau s}^{32}$,\\
			$\mathcal{A}_{\tau s}^{43}=  \mathcal{A}_{\tau s}^{43}$.
			\end{tabular}
		\end{table}

\end{appendix}
\bibliography{article.bib}{}
\bibliographystyle{unsrt}
\end{document}